\documentstyle[prd,floats,aps]{revtex}

\begin{document}
\draft

\input{epsf.sty}

\twocolumn[\hsize\textwidth\columnwidth\hsize\csname
@twocolumnfalse\endcsname

\title{The Inflaton Field as 
Self-Interacting Dark Matter in the Braneworld Scenario}

\author{James E. Lidsey$^{1}$, Tonatiuh~Matos$^{2}$ and
L. Arturo~Ure\~{n}a-L\'{o}pez$^{2}$ }

\address{$^{1}$Astronomy Unit, School of Mathematical 
Sciences, Queen Mary, University of London,  Mile End 
Road, London E1 4NS, United Kingdom.\\
$^{2}$Departamento de F\'{\i}sica, Centro 
de Investigaci\'{o}n y de Estudios Avanzados 
del IPN, A.P. 14-740, 07000 M\'{e}xico D.F., M\'{e}xico.}

\date{\today}

\maketitle

\begin{abstract}
A unified model 
is developed within the context of the 
braneworld paradigm, where 
a single scalar field can act as both the inflaton  
field in the very early universe and also as strong, 
self--interacting 
dark matter in the post--inflationary universe.
Reheating proceeds due to the overproduction and 
subsequent evaporation of primordial black holes. 
Observational constraints, most notably from gravitational 
waves, are satisfied if the probability of PBH formation 
is sufficiently high.

\end{abstract}


\vskip2pc]

\narrowtext   
The recent advances in astronomical observations are casting new light on
old problems in cosmology. Two of the most fundamental questions of interest
today are: (i) the origin of the scalar inflaton field responsible for the
accelerated, inflationary expansion of the very early universe and: (ii) the
nature of dark matter in the universe. Both questions have important
implications for our understanding of large--scale structure formation \cite
{ll}.

New ideas about inflationary cosmology are now emerging with the development
of the braneworld scenario, where our observable universe is viewed as a
domain wall embedded within a higher--dimensional space \cite{brane,RS2}. A
striking feature of this scenario is the presence of a quadratic density
term in the Friedmann equation \cite{quadratic}. Under quite general
conditions, this term allows steep scalar field potentials to support an
inflationary epoch that would otherwise be impossible in standard cosmology 
\cite{maartens,lidsey,lidsey1,varun}.

An attractive feature of steep inflationary models is that the universe can
be naturally reheated by the process of gravitational particle production,
where particles are produced quantum mechanically at the end of inflation
due to the time--variation of the gravitational field \cite{gpp}. 
If the inflaton is stable and is able to survive through to the 
present epoch, it may represent a possible candidate for the 
cosmological constant or quintessence field 
\cite{BK1,earlier,lidsey,lidsey1,varun} 
that has been proposed to account for the high
redshift type Ia supernova data \cite{quintessence,sn}.

On the other hand, Barshay and Kreyerhoff first proposed that a metastable 
inflaton field 
could be identified with the present--day cold dark matter (CDM)
\cite{BK1}. 
By employing renormalization--group techniques to calculate radiative 
corrections to a scalar field potential involving a quartic self--interaction 
\cite{BK2}, 
these authors found that inflation could proceed in the vicinity of 
the potential's maximum and would end as the field reached a well--defined 
minimum. Fluctuations about this minimum result in massive inflaton quanta 
with an energy density that is sufficiently high that they 
can provide a significant 
fraction of the CDM at the present epoch. 

In this paper, we 
consider whether the inflaton field can be identified with the cold dark 
matter in the universe \cite{BK1}, within the specific context of the 
braneworld scenario and the so--called 
`strong, self--interacting scalar field dark matter' (SFDM) hypothesis 
that has recently been developed by two of the authors  
\cite{QSDMCQG,COSPRD,franky,cross,luis,futuro,futuro1} (see also\cite{sahni}). The key idea of the SFDM scenario is that the dark matter responsible for 
structure formation in the universe is a real scalar 
field, $\Phi$, that is minimally coupled to Einstein 
gravity and has self-interactions parametrized by a potential energy 
of the form 
\begin{equation}
V(\Phi )=V_{0}\left[ \cosh (\alpha \sqrt{\kappa _{0}}\Phi )-1 \right] \,,
\label{coshpot}
\end{equation}
where $V_{0}$ and $\alpha $ are the only two free parameters of the model, $%
\kappa _{0}=8\pi G$ and we employ natural units such that $\hbar =c=1$. The
effective mass of the scalar field is given by $m_{\Phi }^{2}=\kappa
_{0}V_{0}\alpha ^{2}$. A minimal coupling to 
gravity avoids the strong restrictions
imposed by the equivalence principle on scales of the order of the solar
system.

The advantage of the SFDM model is that it is insensitive to initial
conditions and the field behaves as CDM once it begins to oscillate around
the minimum of its potential. In this case, it can be shown \cite
{QSDMCQG,COSPRD} that the SFDM model is able to reproduce all the successes
of the standard $\Lambda $CDM model above galactic scales. Furthermore, it
predicts a sharp cut-off in the mass power spectrum due to its quadratic
nature, thus explaining the observed dearth of dwarf galaxies, in contrast
with the excess predicted by high resolution N-body simulations with
standard CDM \cite{COSPRD,dmp}. The strong self-interaction of the scalar field
results in the formation of solitonic objects called `oscillatons', which
have a mass of the order of a galaxy but do not exhibit the cusp density
profiles characteristic of standard CDM \cite{luis,futuro,futuro1}. The
best--fit model to the cosmological data can be deduced from the current
densities of dark matter and radiation in the universe and from the cut--off
in the mass power spectrum that constrains the number of dwarf galaxies in
clusters. The favoured values for the two free parameters of the potential (%
\ref{coshpot}) are found to be \cite{COSPRD}: 
\begin{eqnarray}
\alpha &\simeq &20.28\,,  \label{lambda} \\
V_{0} &\simeq &(3\times 10^{-27}\,M_{4})^{4}\,,  \label{V0}
\end{eqnarray}
where $M_{4}\equiv G^{-1/2} \approx 10^{-5}$g is the four--dimensional
Planck mass. This implies that the effective mass of the scalar field should
be $m_{\Phi }\simeq 9.1\times 10^{-52}\,M_{4}=1.1\times 10^{-23}$ eV.

An important feature of the potential (\ref{coshpot}) is that it is
renormalizable and exactly quantizable, although 
it is presently unknown whether it
originates from a fundamental quantum field theory 
\cite{halpern}. 
Furthermore, the
scattering cross section by mass of the scalar particles, $\sigma
_{2\rightarrow 2}/m_{\Phi }$, can be constrained from numerical simulations
of self-interacting dark matter that avoid high-density dark matter 
halos \cite{sdm}.
This effectively constrains the renormalization scale, $\Lambda_{\Phi}$, of
the potential to be of the order of the Planck mass, $\Lambda _{\Phi }\simeq
1.93M_{4}=2.15\times 10^{19}$ GeV \cite{cross}. Such a value is indicative
of a possible fundamental origin for the scalar field, which in turn
suggests that the strongly, self--interacting scalar field dark matter may
also have been present during the inflationary epoch \cite{cross}.
However, we do not perform quantum field theoretic or semi--classical 
calculations in this paper. 

In view of the high energy scales associated with the very early universe,
it is natural to assume that the scalar field was initially displaced from
its global minimum, $\kappa _{0} \alpha^2 \Phi ^{2}\gg 1$. In this limit,
the potential is well approximated by an exponential function, but from Eq. (%
\ref{lambda}), the self--coupling is too large to support inflationary
expansion in a standard cosmological setting. On the other hand, such a
potential can drive inflation successfully within the braneworld scenario 
\cite{lidsey,varun}. This follows because the Friedmann equation is modified
due to the motion of our observable universe (the domain wall) through the
higher--dimensional `bulk' spacetime. In particular, in the second
Randall--Sundrum scenario \cite{RS2}, the Friedmann equation is given by 
\cite{quadratic} 
\begin{equation}
H^{2}=\frac{\kappa_0}{3}\rho \left( 1+\frac{\rho }{2\lambda _{b}} \right)
\label{friedmann}
\end{equation}
when appropriate conditions are satisfied, where $H\equiv \dot{a}/{a}$ is
the Hubble parameter, $a$ is the scale factor of the universe, $\rho $ is
the energy--density of the inflaton field (assumed to be confined to the
brane) and $\lambda _{b}$ is the tension of the brane. Conventional Einstein
gravity is recovered in four dimensions when the energy density is
significantly lower than the brane tension, $\rho \ll \lambda _{b}$.
However, at high energies, the quadratic correction implies that the
expansion rate of the brane is enhanced relative to what it would be in a
universe governed by Einstein gravity \cite{maartens}. Thus, the friction
acting on the scalar field is increased and inflation driven by a potential
of the form (\ref{coshpot}) is then possible at sufficiently high energies
even though $\alpha ^{2}\gg 1$.

Braneworld inflation driven by such a potential has been studied in Refs. 
\cite{lidsey,varun}. Recalling the main results, the COBE normalization of
the cosmic microwave background (CMB) power spectrum \cite{cobe} relates the
value of the brane tension to the scalar field self--coupling such that $
\lambda _{b}^{1/4}\alpha ^{3/2}\approx 10^{15}$ GeV. For the favoured value
of the latter, as implied by Eq. (\ref{lambda}), we deduce that 
\begin{equation}
\lambda _{b}\simeq \left( 6\times 10^{-7}M_{4}\right) ^{4}=2.88\times
10^{51}\, {\rm GeV}^{4} .  \label{lambdab}
\end{equation}
For these given values of $\{\alpha ,\lambda _{b}\}$, the magnitude of the
potential energy at the end of inflation is $V_{{\rm end}}\simeq (3.2\times
10^{-6}M_{4})^{4}=2.33\times 10^{54}\, {\rm GeV}^{4}$ and this implies that $
\Phi_{{\rm end}}\approx 2M_{4}$, thus justifying the exponential
approximation to the potential (\ref{coshpot}) during the inflationary era.
In comparison with the canonical $V= \lambda \Phi^4$ potential, where 
COBE normalization implies the dimensionless 
parameter $\lambda \approx 10^{-13}$ should be very small, 
our model requires the small dimensionful number for $V_0$ given in 
Eq. (\ref{V0}), although this constraint does not arise from the CMB. 
Furthermore, the magnitude of the potential $N$ $e$--foldings before 
the end of inflation is $V_N \approx (N+1)V_{\rm end}$, implying that 
the initial value of the scalar field is closer to $\Phi_{\rm end}$ than 
in the quartic model. 

Given the COBE normalization \cite{cobe}, the spectral index of the scalar
fluctuation spectrum is determined to be \cite{lidsey} 
\begin{equation}  \label{tilt}
n=1-\frac{4}{N+1}=0.94,
\end{equation}
where $N \approx 70 $ is the number of {\it e}-foldings that elapse between
the epoch that a given, observable mode crosses the Hubble radius during
inflation and the end of the inflationary epoch. Remarkably, the tilt of the
scalar perturbation spectrum in this scenario is {\em uniquely} determined
by the number of {\em e}--foldings and is {\em independent} of the
parameters in the potential (\ref{coshpot}). A spectrum with a tilt of this
magnitude away from scale--invariance is presently favoured by analyses of
the CMB power spectrum \cite{teg}. Furthermore, the amplitude of the
primordial gravitational wave spectrum, $A_T$, relative to that of the
density perturbations, $A_S$, can be estimated as \cite{lidsey} 
\begin{equation}  \label{ratio}
r=4\pi \frac{A_{T}^{2}}{A_{S}^{2}}\simeq 0.4
\end{equation}
implying after COBE normalization that $A_{T}^{2}\approx 1.7\times 10^{-10}$%
. This ratio is also independent of the model's parameters and is within the
projected sensitivity of the Planck satellite. It provides a potentially
powerful test of the model.

Inflation ends when the quadratic corrections to the Friedmann equation (\ref
{friedmann}) become negligible. Due to the steep nature of its potential,
the inflaton then behaves as a massless field, where its energy density
redshifts as $\rho_{\Phi} \propto a^{-6}$. This is important, because after
the tensor modes have re--entered the Hubble radius, the evolution of the
spectral gravity wave energy density, $\tilde{\rho}_g$, is sensitive to the
effective equation of state in the post--inflationary universe\footnote{%
The equation of state is assumed to be of the form $p = \omega \rho$, where $%
\omega$ is a constant barotropic index. For a massless scalar field, $\omega
=1$.} \cite{varun,allen}. It is enhanced (reduced) on shorter scales if $%
\omega > 1/3$ $(\omega < 1/ 3)$. In general, the bound on the gravitational
waves imposed by successful nucleosynthesis, $\rho_g \le 0.2 \rho_{{\rm rad}%
} $, must not be violated and this results in an upper limit on the duration
of the kinetic energy dominated phase \cite{varun}.

Sahni, Sami and Souradeep \cite{varun} assume that the evolution of the
short wavelenth gravitational waves is similar to that of conventional
cosmology. They then conclude that the gravitational wave energy density
begins to dominate the scalar field when the universe has expanded by a
factor of $A_T^{-1} \approx 10^5$ and, consequently, the universe must
become radiation dominated before the temperature has fallen by this factor.
Unfortunately, for the model under consideration, the thermalized
temperature of the radiation produced from gravitational particle production
at the end of inflation is given by $T_{{\rm end}} \simeq 2 \times 10^{9}$
GeV, whereas the temperature at the epoch when this radiation dominates the
scalar field is $T_{{\rm eq}} \simeq (1-2)$ GeV \cite{varun}, corresponding
to a redshift of $10^9$.

An alternative mechanism for reheating the universe is therefore required
that reduces the duration of the kinetic energy dominated phase. One
possibility is through the overproduction of primordial black holes (PBHs)
that subsequently decay into relativistic particles via Hawking evaporation 
\cite{hawking,cl,heatpbh}. In the above inflationary model, a fraction, $%
\beta_0$, of the energy density of the universe collapses into PBHs due to
the density fluctuations that re--enter the Hubble radius immediately after
the universe has ceased to accelerate. The PBHs form with a mass of the
order of the horizon mass at this time \cite{carr} and this is given by $M_{%
{\rm pbh}} \approx M^2_4H^{-1}_{{\rm end}}$, where the Hubble radius at the
end of inflation, $H^{-1}_{{\rm end}}$, is estimated from the Friedmann
equation (\ref{friedmann}) under the assumption that $\rho_{{\rm end}}
\approx V_{{\rm end}} \approx 2 \alpha^2 \lambda_b$. Eq. (\ref{lambdab})
then implies that $M_{{\rm pbh}} \approx 10^9 M_4$ and the lifetime of PBHs
with this mass is $t_{{\rm evap}} \approx (M_{{\rm pbh}} /M_4)^3t_P \approx
10^{-16}$s, where $t_P$ is the Planck time. This is sufficiently short that
constraints on PBH evaporations from primordial nucleosynthesis are
satisfied \cite{cl,gl}.

Once formed, the PBHs behave as a pressureless fluid and their energy
density redshifts as $\rho_{{\rm pbh}} \propto a^{-3}$. Thus, they are able
to dominate the (massless) scalar field before they evaporate if 
\begin{equation}  \label{dominate}
\frac{\beta_0}{1-\beta_0} > \left( \frac{M_{{\rm pbh}}}{M_4} \right)^{-2}
\approx 10^{-18}
\end{equation}
and this change in the effective equation of state occurs after the universe
has expanded by a factor 
\begin{equation}  \label{before}
\frac{\beta_0}{1- \beta_0} \approx \left( \frac{a_{{\rm end}}}{a_{{\rm dom}}}
\right)^3,
\end{equation}
where $a_{{\rm dom}}$ denotes the scale factor at the onset of PBH
domination. It follows, therefore, that the PBHs dominate the scalar field
before the backreaction of the gravitational waves becomes significant if
the initial mass fraction satisfies $\beta_0 > 10^{-15}$. This is consistent
with Eq. (\ref{dominate}) and the constraint arising from the integrated
gravitational wave energy density is therefore alleviated since the PBH
equation of state, $\omega =0$, is `softer' than that of radiation.

A stronger, and more reliable, constraint on the initial PBH fraction can be
imposed by requiring that the PBHs dominate the universe before the scalar
field has reached the minimum of its potential and begun to oscillate.
During the kinetic dominated regime, the scalar field varies as 
\begin{equation}
\Phi = \Phi_{{\rm end}} -\sqrt{\frac{3}{4\pi}} M_4 \ln \left( \frac{a}{a_{%
{\rm end}}}\right)
\end{equation}
and from the estimate of $\Phi _{{\rm end}}$ given above, $\Phi _{{\rm end}%
}\approx 2M_{4}$, the field reaches the minimum of its potential after the
universe has redshifted by a factor $a/a_{{\rm end}}\approx 60$. Thus, from
Eq. (\ref{before}), the PBHs dominate the universe before this point is
reached if $\beta _{0}>5 \times 10^{-6}$.

If the PBHs come to dominate sufficiently early, the displacement of the
scalar field away from its minimum is such that the potential is still well
approximated by an exponential form at this time. This is the case for $\Phi
>0.01M_{4}$. Moreover, the standard Friedmann equation is valid for $\Phi
<1.8M_{4}$, where the quadratic corrections in Eq. (\ref{friedmann}) become
negligible. Since its potential is steep, the field subsequently tracks the
PBH (fluid) component in this regime as in the standard cosmology \cite
{stantrack}, where its potential and kinetic energies scale at the same rate
as that of the PBH energy density. More specifically, $\Omega _{\Phi
}=3(1+\omega )/\alpha ^{2} $ and $\dot{\Phi}^{2}/V=2(1+\omega )/(1-\omega )$
and this implies that the variation of the scalar field with respect to the
scale factor during tracking is given by \cite{stantrack} 
\begin{equation}
\Phi =\Phi_{{\rm t0}} -\frac{3(1+\omega )}{\sqrt{\kappa _{0}}\alpha }\ln
\left( \frac{a}{a_{{\rm t0}}}\right) ,  \label{tracking}
\end{equation}
where a subscript `t0' denotes the onset of the tracking behaviour. This is
important because the large value of the field's self--coupling, $\alpha
\approx 20$, implies that the universe expands by many orders of magnitude
before the field reaches its global minimum. In particular, if the PBHs
dominate the cosmic dynamics for the majority of their lifetime, the
universe can expand by up to a factor of $10^{12}$ before the PBHs
evaporate. The rolling of the scalar field down its potential during this
epoch is only $\Delta \Phi \approx -0.84$. The subsequent transition to a
radiation dominated universe has a negligible effect on the tracking
behaviour of the field, and for a wide range of initial conditions, the
field does not reach its minimum until well after the primordial
nucleosynthesis era has passed. In this case, the nucleosynthesis bounds are
not violated since $\alpha >5$ \cite{pedro}. In principle, therefore, our
model does not exhibit the problem of overshooting nor undershooting for the
initial conditions \cite{pedro}. The post--inflationary universe after the
nucleosynthesis era would then correspond to the universe considered in
Refs. \cite{QSDMCQG,COSPRD,franky,cross}, where it was shown that the scalar
field can subsequently act as dark matter in the universe.

To summarize, we have found that a scalar field with a potential of the form
(\ref{coshpot}) can drive an epoch of inflationary expansion in the
braneworld scenario and may also act as a candidate for the dark matter at
the present epoch. The reheating of the universe proceeds via PBH domination
and evaporation. The model is able to explain a variety of astrophysical
observations over a wide range of scales and, indeed, contains only four,
free parameters: the effective mass of the scalar field, $m_{\Phi}$, its
self--coupling, $\alpha$, the tension of the brane, $\lambda_b$, and the
initial mass fraction of PBHs, $\beta_0$. It is the insensitivity of Eqs. (%
\ref{tilt}) and (\ref{ratio}) to the potential's parameters that allows us
to simultaneously satisfy the constraints arising from the mass power
spectrum without the need for fine--tuning. Once fixed by observational
constraints, the parameters of the potential can remain unaltered throughout
the history of the universe.

In this paper, we have considered the potential (\ref{coshpot}) 
from a phenomenological perspective with the primary aim 
of determining the region of parameter space consistent with 
astrophysical observation. The parameters cover a wide range of 
scales and 
one question that immediately arises is whether the form of the potential 
(\ref{coshpot}) and, in particular, the best--fit values 
for its parameters, can be more fully understood from a field--theoretic 
perspective. This is beyond the scope of the present paper. 
Nevertheless, 
we have developed a model based on a single scalar field that 
covers the history of the universe from the inflationary era
through to the present epoch and 
it is therefore to be expected that the parameters should cover a similar 
range of energy scales. 
The success of the model in simultaneously 
satisfying a number of fundamental observational 
constraints provides strong motivation for considering 
its field--theoretic origin 
in more detail.

We have viewed the initial mass fraction of PBHs as a free parameter in the
above analysis, but its magnitude is determined by the density perturbations
at the end of inflation. A more detailed calculation of the fluctuation
spectra is therefore required if further insight is to be gained, but this
involves an extension of the slow--roll analyses employed thus far \cite
{maartens} and is beyond the scope of the present paper. Since the
post--inflationary universe is initially dominated by a scalar field, the
question of PBH formation in this scenario is closely related to the problem
of scalar field collapse and this topic has attracted considerable attention
in recent years. (For a recent review, see, e.g., Ref. \cite{recent}).
Moreover, in estimating the limits on the probability of PBH formation, we
have assumed that standard four--dimensional results derived within the
context of Einstein gravity are valid. This is reasonable since the PBHs
form once inflation has ended and this occurs when the brane corrections to
the Friedmann equation (\ref{friedmann}) have become negligible.

A potential problem with reheating a braneworld inflationary universe via
PBH evaporations is that the decay products may radiate primarily off the
brane and into the higher--dimensional bulk, thereby rendering the brane
effectively cold and empty. This remains an open question in the second
Randall--Sundrum scenario, but it has been shown in a related model that
most of the Hawking radiation consists of standard model particles that are
indeed confined to the brane \cite{myers}.

Finally, there is the question of whether the PBHs leave behind stable,
Planck--sized relics at the endpoint of their evaporation \cite{cl,bcl}.
Although such a possibility is now considered unlikely, it is worth
remarking that when the PBHs dominate the universe before they evaporate,
their relics can not overclose the universe if $M_{{\rm pbh}} < 10^6$g \cite
{cl} and this bound is satisfied for the above scenario.

In conclusion, we have proposed a simple, unified model of the inflaton and
dark matter particles, where the same scalar field provides the origin for
the primordial spectrum of density perturbations produced quantum
mechanically during inflation and also plays a central role in forming the
cosmological structures that we observe today.


\acknowledgements{JEL is supported by the 
Royal Society. LAUL wants to thank the Theoretisch-Physikalisches
Institut (Jena) for its kind hospitality. TM and LAUL are partly supported 
by CONACyT M\'exico, under 
grants 119259 (L.A.U.) and 34407-E. JEL's visit to CINVESTAV was 
supported by CONACyT grant 38495-E. We thank C. Terrero--Escalante for 
a helpful discussion.}

\end{document}